\documentclass[twocolumn,pre]{revtex4}
\usepackage{graphicx}
\usepackage{dcolumn}
\usepackage{bm}
\usepackage{amsmath,amssymb}

\newcommand {\pdd}[2]{\frac{\partial #1}{\partial #2}}

\newcommand {\vel}{\mathbf{u}}

\begin{document}

\title{Evaporatively driven morphological instability}

\author{Robert W. Style$^{1}$ and J. S.~Wettlaufer$^{2}$ }
\affiliation{$^1$Institute of Theoretical Geophysics, Department of Applied Mathematics and Theoretical Physics, University of Cambridge, Wilberforce Road, Cambridge, CB3 0WA,  U.K.}
\affiliation{$^2$Department of Geology \& Geophysics and Department of Physics, Yale University,
 New Haven, Connecticut 06520-8109, USA}

\begin{abstract}
Simple observations of evaporating solutions reveal a complex hierarchy of spatio-temporal instabilities.  We analyze one such instability suggested by the qualitative observations of Du and Stone and find that it is driven by a novel variant of the classical {\em morphological instability} in alloy solidification.  In the latter case a moving solid--liquid interface is accompanied by a solutally enriched boundary layer  that is thermodynamically metastable due to {\em constitutional supercooling}.  Here, we consider the evaporation of an impure film adjacent to a solid composed of the nonvolatile species.  In this case, constitutional supercooling within the film is created by evaporation at the solution--vapor interface and this drives the corrugation of the solid--solution interface across the thickness of the film.  The principal points of this simple theoretical study are to suggest an instability mechanism that is likely operative across a broad range of technological and natural systems and to focus future quantitative experimental searches.   
\end{abstract}

\date{23 June, 2007}

\maketitle

\section{Introduction and Motivation}

Evaporation of water is an important phase transformation appearing in many guises throughout everyday life. Its effects range from processes intrinsically linked to the Earth's hydrologic cycle, to the regulation of body temperature in hot environments, to the production of coffee--ring stains beneath a spilt coffee droplet and to important processes underlying microfluidics.  
While the evaporation of bulk pure materials provides a wide variety of thin film instabilities \cite{oron97}, the simple act of introducing solutes, with their myriad of interfacial effects, leads to a spate of new behaviors whose exploration is in its infancy.  Examples include the investigation of coffee--ring formation by a sessile, particle--laden droplet \cite{deeg00}, the observation of finite contact angles in evaporating, wetting films \cite{elba94,lips96}, and of particular interest here, the observations of Du \& Stone on evaporatively grown salt trees \cite{du96}.  Understanding the underlying mechanisms that control the patterns that emerge during evaporation and solidification of solutions influences a wide range of fields, from the interpretation of  {\em evaporite deposits} for clues about environmental conditions in the geologic past \cite{evaporites}, to the {\em scaling} deposits in tea kettles when used with hard water \cite{scaling}.   

\begin{figure}[[!h]
\label{amchlo}
\centering
\includegraphics[height=10cm]{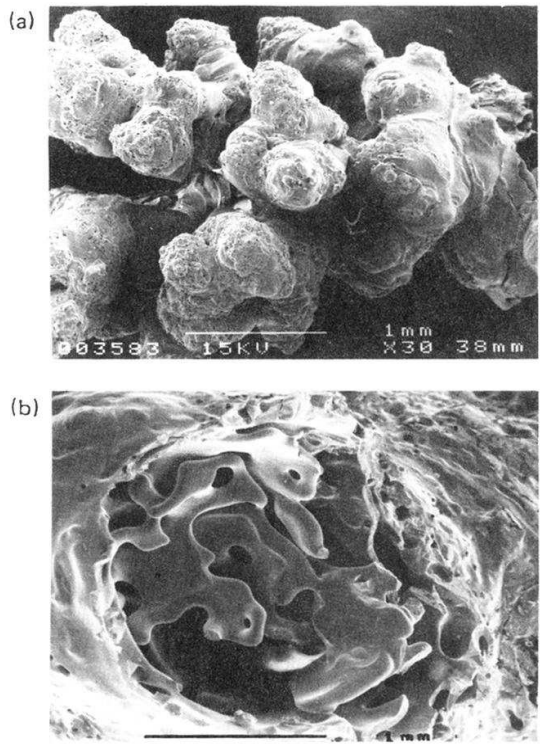}
\caption{Scanning electron micrographs of typical ammonium chloride salt trees. (a) A branch of a dried tree approximately 3mm in length. The picture demonstrates the multiscale roughening resulting from the evaporative process.  (b) Cross section of the porous interior of a typical dried tree. Roughening is caused by evaporation in the presence of a wetting film \cite{du96}. These images motivate our study
but do not provide a length scale against which theory can be quantitatively compared.}
\end{figure}

Du \& Stone \cite{du96} observed the evolution of ramified crystal structures during the evaporation of water from an ammonium chloride solution under varying (but not quantitatively controlled) external conditions. Initially these crystals were confined within the bulk liquid, but eventually they protruded above the liquid surface, forming {\em salt trees}, the growth in this latter stage being caused by evaporation of a film maintained at the surface of the solid salt by capillarity.  In consequence Du \& Stone \cite{du96} postulated the existence of two distinct instabilities giving rise to two different characteristic lengthscales of the system (figure~\ref{amchlo}).   Although their observations did not quantify the conditions for onset of either instability they were led to make the following general distinctions.  The smaller scale instability (with a characteristic length scale estimated to be $\approx$ 10$\mu$m) was suggested to arise from evaporatively driven supersaturation that allows fluctuations of the salt--liquid interface to grow in a manner qualitatively similar to the growth of a solid into a supercooled melt. The second instability (with a characteristic length scale estimated to be $\approx$ 100$\mu$m) was proposed as being due to a spatial variation in evaporative flux across the surface; the film at the surface of a crystal  protruding further into the dry atmosphere will cause faster evaporation of the film and hence further growth relative to adjacent regions.  

While these rough estimates neither constrain the nature of either instability, nor quantify the thermodynamic ranges of their existence, the Du \& Stone \cite{du96} observations motivate a simple theoretical analysis that we hope will focus the future experimental search.  Indeed, more sophisticated treatments are not warranted by the present experimental evidence.  Hence, in this paper we focus on the first of these instabilities and  demonstrate the existence of an observable length-scale during the evaporative growth of salt crystals.  

\section{Theory}

We study the mechanism of the first, salt-diffusion-driven, instability and simplify the system as follows.  Firstly, although latent heat is produced at two phase boundaries, because the diffusivity of heat in the liquid phase $\kappa_w$ is much greater than the diffusivity of salt $D$, we are justified in treating the system as isothermal.  Secondly, the vapor pressure of salt in air is exceedingly small relative to that of water vapor and hence there is no salt transport in the air.  Additionally, apart from the consequences of the instability we discuss, there are no liquid inclusions in the salt phase.
Thirdly, we treat the system as planar with shallow perturbations to the two interfaces.  This allows us to use the typical approximations to the curvature and normal derivatives that we check {\em{post~hoc}}. 
Finally, we assume that the liquid--vapor interface is maintained flat by flow in the liquid phase due to the influence of surface tension. From lubrication theory, we find that the time scale for the relaxation of a film of thickness $d_0$ to a planar state is given by
\begin{equation}
\tau_{\mu}\sim \frac{\mu}{\sigma_{la}\alpha^4d_0^3}
\end{equation}
where $\mu$ is the dynamic viscosity of water, $\alpha$ is the wavenumber of the perturbation, and $\sigma_{la}$ is the surface tension of the liquid--air interface. The diffusive time scale for the system is given by $\tau_D=d_0^2/D$ where $D$ is the diffusivity of salt in the liquid. Thus the hydrodynamic relaxation to a planar interface will occur on a much shorter timescale than that on which instability occurs as long as
\begin{equation}
\frac{\mu\omega_{max}}{\sigma_{la}\alpha^4d_0^3}\ll1,
\end{equation}
where $\omega_{max}$ is the maximum growth rate of the instability as given later.
Hence, because $\alpha\sim 2\pi/10^{-5}$m (as we show later), we can ignore the motion of the liquid--vapor interface when $d_0\gg6\times10^{-10}$m.  This is a good assumption in the relevant parameter space. We will also show that the advection of solute in the film is negligible relative to diffusive transport, and so the precise details of the flow can be ignored.
These are either common approximations that are borne out experimentally or are confirmed to hold within the regimes of validity of the theory. Finally, although we have analysed the effect of marangoni stresses in transporting liquid to the edges of evaporating droplets \cite{wett99c} previously, the salt concentration at the liquid--vapor interface was taken as fixed, giving rise to large salinity differences across the liquid film. Here, we relax this assumption and consider instead the evaporatively controlled salinity at the liquid--vapor interface.  Motivated by the observations of Du \& Stone \cite{du96}, we shall treat the ammonium chloride--water system at room temperature and typical humidities throughout this paper. However, we note that the analysis is applicable to many other systems of the same general class.  

\begin{figure}[!h]
\centering
\includegraphics[height=8cm]{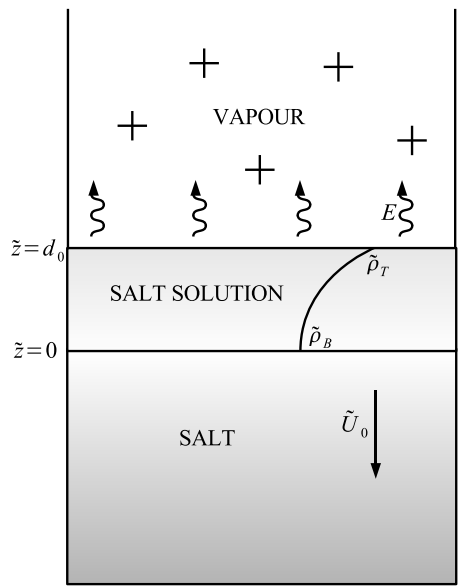}
\caption{Schematic for the planar growth of a salt crystal due to the evaporation of a salty film. $\tilde{\rho_B}$ and $\tilde{\rho_T}$ are the densities of salt in the solution at the salt--liquid interface and the liquid--vapour interface respectively, $E$ is the evaporation rate of the film, and $\tilde{U}_0$ is the growth rate of the salt crystal. Initially the interfacial positions are given by $\tilde{z}=0$ for the salt--liquid interface and $\tilde{z}=d_0$ for the liquid--vapour interface.}
\label{wett}
\end{figure}

Consider a planar salt crystal covered by a film of salt solution that has an initial thickness $d_0$ as shown in figure (\ref{wett}). We expose the film to an undersaturated vapor which drives evaporation at the liquid--vapor interface.  In general, within the film the solute density then satisfies the advection--diffusion equation
\begin{equation}
\label{diffadv}
\pdd{\tilde{\rho}}{\tilde{t}}+\tilde{\vel}.\tilde{\nabla} \tilde{\rho}=D\tilde{\nabla}^2 \tilde{\rho}
\end{equation}
where $\tilde{\rho}(\tilde{x},\tilde{z},\tilde{t})$ is the density of solute in the liquid, $\tilde{\vel}$ is the velocity of the liquid and $\tilde{x},\tilde{z},\tilde{t}$ are horizontal, vertical and time coordinates.

Two-phase, two-component equilibrium at the two phase boundaries, requires the satisfaction of two boundary conditions for the system. Firstly, the liquidus relationship, which defines bulk equilibrium between the salt in the film and that in the underlying planar crystal,  is modified for the shift in equilibrium associated with the curvature of the solid--liquid interface $\tilde{{\cal K}}_{sl}$ as
\begin{equation}
T=T_0+m\tilde{\rho}+\frac{\sigma_{sl}T_m}{{\cal L}_f}\tilde{{\cal K}}_{sl},
\end{equation}
where $T_m$ is the melting temperature of a pure liquid salt, $m$ is the slope of the liquidus, $\sigma_{sl}$ is the surface energy of the salt--liquid interface and ${\cal L}_f$ is the heat of solution of the salt.  This describes equilibrium as a linear combination of the colligative and Gibbs-Thomson effects.  We note that we have assumed that the curvature term is independent of solute density, so the resulting expression will not be exact. Rearranging, we obtain
\begin{equation}
\tilde{\rho}=\frac{T_m-T}{m}+\frac{\sigma_{sl}T_m}{m{\cal L}_f}\tilde{{\cal K}}_{sl}
\end{equation}
which gives us the equilibrium boundary condition at the salt--liquid interface:
\begin{equation}
\label{bc3}
\tilde{\rho}(\tilde{h}_{sl})=\tilde{\rho}_B^0+\gamma \tilde{{\cal K}}_{sl}.
\end{equation}
$\tilde{h}_{sl}$ is the vertical position of the solid--liquid interface relative to the position of the planar growth interface position as derived below, $\tilde{\rho}_B^0=(T_m-T)/m$ is the equilibrium value of the solute density at a planar interface, and $\gamma=\sigma_{sl}T_m/m{\cal L}_f\approx1.8\times10^{-7}\mathrm{kg}\,\mathrm{m}^{-2}$.

Secondly, the dynamically maintained planarity of the liquid--vapor interface allows us to 
treat the evaporation rate there as a constant.  Although the evaporation rate may be influenced by gradients in salinity along the surface, these are shown below to be small because $\tilde{\rho}\approx\tilde{\rho}_B^0$. Thus we take
\begin{equation}
-\dot{\tilde{h}}_{lv}=E.
\end{equation}

The two boundary conditions needed to close the system of equations express conservation of salt across the solid--liquid interface,
\begin{equation}
\label{bc1}
D\left.\pdd{\tilde{\rho}}{\tilde{z}}\right|_{\tilde{z}=\tilde{h}_{sl}}=(\dot{\tilde{h}}_{lv}+\tilde{U}_0)\tilde{\rho}_{salt},
\end{equation}
and conservation of water across the liquid--vapor interface,
\begin{equation}
\label{bc2}
D\left.\pdd{\tilde{\rho}}{\tilde{z}}\right|_{\tilde{z}=\tilde{h}_{sl}}=E\left.\rho\right|_{\tilde{z}=\tilde{h}_{sl}}.
\end{equation}
Here $\tilde{\rho}_{salt}$ is the density of salt in the solid phase and $\tilde{U}_0$ is the planar growth rate of the solid--liquid interface before the introduction of any perturbation.

The natural length, velocity and time scales imposed on the system are given by $d_0$, $E$ and $d_0/E$ respectively and so we scale variables accordingly so that $(\tilde{x},\tilde{z},\tilde{h}_{lv},\tilde{h}_{sl})=d_0(x,z,h_{lv},h_{sl})$, $(\tilde{\vel},\tilde{U}_0)=E(\vel,U_0)$ and $\tilde{t}=d_0t/E$. Thus, equation (\ref{diffadv}) becomes
\begin{equation}
\mathrm{Pe}\left(\pdd{\tilde{\rho}}{t}+\vel.\nabla \tilde{\rho}\right)=\nabla^2 \tilde{\rho},
\end{equation}
where the Peclet number Pe is defined as $d_0 E/D$ and corresponds to the relative importance of advective (due to evaporation of the interface) and diffusive transport processes. For a typical aqueous ammonium chloride solution at $20^{\circ}$C, $E\sim10^{-7}$m$\,\mathrm{s}^{-1}$ and $D=10^{-9}\mathrm{m}^2\, \mathrm{s}^{-1}$ so that Pe$\sim d_0(\mathrm{m})\times10^2$.  Therefore, the Peclet number will be small for all films under consideration and we can approximate equation (\ref{diffadv}) by Laplace's equation
\begin{equation}
\nabla^2\tilde{\rho}=0.
\end{equation}
Hence, during planar growth of the salt crystal information is diffused sufficiently rapidly across the film that the solutal profile is linear and given by
\begin{equation}
\tilde{\rho}=(\tilde{\rho}_T^0-\tilde{\rho}_B^0)z+\tilde{\rho}_B^0.
\end{equation}
Here, $\tilde{\rho}_T^0$ is the solute density at the liquid--vapor interface in this basic state, 
determined from the boundary conditions (\ref{bc1}) and (\ref{bc2}) upon application of which we find
\begin{equation}
\label{a}
\tilde{\rho}_T^0=\tilde{\rho}_B^0\left(\frac{1}{1-\mathrm{Pe}}\right),
\end{equation}
and
\begin{equation}
\label{b}
\tilde{U}_0\tilde{\rho}_{salt}=E\tilde{\rho}_T^0.
\end{equation}
We note that this last equation agrees with our intuition expressing as it does the fact that as water is evaporated at the surface, the amount of salt that was dissolved in the evaporated water is instantaneously deposited onto the salt crystal at the base of the film. Also, from equation \ref{a} the gradient in salt density across the film is given by
\begin{equation}
\frac{\tilde{\rho}_T^0-\tilde{\rho}_B^0}{d_0}=\frac{\tilde{\rho}_T^0\mathrm{Pe}}{d_0}\approx\frac{\tilde{\rho}_B^0E}{D}.
\end{equation}
The last approximation also stems from the fact that the Peclet number is small. This shows that the salt density gradient across the film is roughly independent of film thickness, and thus any perturbation at the salt--liquid interface will encounter similar salt field conditions independent of film thickness. Thus, in this limit, it can be anticipated that the growth rate of the perturbation will not depend on $d_0$.

We are now in a position to conduct a linear stability analysis of this quasi--stationary planar state, noting first that for this quasi--stationary linear stability analysis to be relevant, the perturbation must grow much faster than the rate of change of the underlying base state, so we must check {\emph{a posteriori}} that the growth rate of the most unstable wavelength satisfies $\sigma_{max}>\tilde{\rho}^0_T/\tilde{\rho}_{salt}$. From consideration of the basic state, we choose to nondimensionalise the solute density by setting $\tilde{\rho}-\tilde{\rho}_B^0=(\tilde{\rho}_T^0-\tilde{\rho}_B^0)\rho$. The governing equations (\ref{diffadv},\ref{bc3},\ref{bc1},\ref{bc2}) then become
\begin{equation}
\nabla^2\rho=0,
\end{equation}
\begin{equation}
\label{1}
\rho=\Gamma{\cal K}_{sl}
\end{equation}
\begin{equation}
\label{2}
\left.\pdd{\rho}{z}\right|_{z=h_{sl}}=\frac{\tilde{\rho}_{salt}}{\tilde{\rho}_T^0}\dot{h}_{sl}
\end{equation}
and
\begin{equation}
\label{3}
\left.\pdd{\rho}{z}\right|_{z=h_{lv}}=\mathrm{Pe}\left.\rho\right|_{z=h_{lv}}+\frac{\tilde{\rho}_B^0}{\tilde{\rho}_T^0}.
\end{equation}
Here, $\Gamma=\gamma/(\tilde{\rho}_T^0-\tilde{\rho}_B^0)d_0$ is the nondimensional solutal Gibbs--Thomson coefficient, and we recall that the interface position $h_{lv}=1-t$. We now impose a small perturbation upon the interface so that
\begin{equation}
h_{sl}=\epsilon e^{i\alpha x+\sigma t}.
\end{equation}

\begin{figure}[!h]
\centering
\includegraphics[height=5cm]{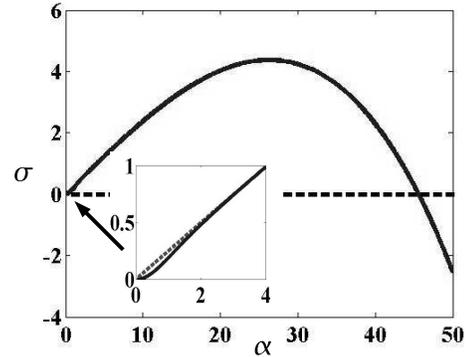}
\label{wett1}
\caption{The dispersion relation for the instability with $d_0=10^{-4}$m. Insert shows the behaviour close to the origin. Continuous line gives full dispersion, dashed line shows large $\alpha$ asymptotic relation.}
\end{figure}

The perturbation to the solutal field that satisfies Laplace's equation is given by
\begin{equation}
\rho=z+(A\sinh\alpha z+B\cosh \alpha z)e^{i\alpha x+\sigma t},
\end{equation}
and after applying boundary conditions (\ref{1}-\ref{3}), we obtain 
\begin{equation}
\alpha A=\epsilon\frac{\tilde{\rho}_{salt}}{\tilde{\rho}_T^0}\sigma,
\end{equation}
\begin{equation}
B=\epsilon(\Gamma\alpha^2-1)
\end{equation}
and
\begin{equation}
A\sinh\alpha+B\cosh\alpha=\frac{A\alpha\cosh\alpha}{\mathrm{Pe}}+\frac{B\alpha\sinh\alpha}{\mathrm{Pe}}
\end{equation}
from which we obtain the dispersion relation governing the instability:
\begin{equation}
\sigma=\frac{\tilde{\rho}_T^0}{\tilde{\rho}_{salt}}\alpha(1-\Gamma\alpha^2)\left(\frac{\alpha\tanh \alpha-\mathrm{Pe}}{\alpha-\mathrm{Pe}\tanh \alpha}\right).
\end{equation}
This relation provides the most unstable wavenumber or cutoff wavelength as
\begin{equation}
\alpha_{co}=\frac{1}{\sqrt{\Gamma}},
\end{equation}
which for a film thickness of $d_0=10^{-4}$m takes a value of $\alpha_{co}=45.3$. Due to the fact that this is much greater than 1, we suggest that a reasonable approximation to the dispersion relation is to take the limit for large $\alpha$, in which case it reduces to the same form as that of the Mullins--Sekerka instability (e.g., \cite{wors00}), namely, 
\begin{equation}
\label{asymp}
\sigma=\frac{\tilde{\rho}_T^0}{\tilde{\rho}_{salt}}\alpha(1-\Gamma\alpha^2).
\end{equation}
Figure (\ref{wett1}) shows a plot of the dispersion relation for $d_0=10^{-4}$m and also the asymptotic approximation given by the equation above. It can be seen that the asymptotic approximation gives excellent agreement with the full expression. Indeed for film thicknesses greater than a micron, the asymptotic expression can be substituted for the full expression without appreciable loss of accuracy.

\begin{table}
\caption{Table of typical values for the ammonium chloride/water system}
\label{tov}
\begin{center}
\begin{tabular}{|l|r|r|}
\hline
Constant & Value & Units \\
\hline
$D$ & $10^{-9}$ & $\mathrm{m}^2\,\mathrm{s}^{-1}$ \\
$E$ & $10^{-7}$ & $\mathrm{m}\,\mathrm{s}^{-1}$ \\
$\gamma$ & $1.8\times10^{-7}$ & $\mathrm{kg}\,\mathrm{m}^{-2}$ \\
$\tilde{\rho}_{salt}$ & $1.5\times10^{3}$ & $\mathrm{kg}\,\mathrm{m}^{-3}$ \\
$\tilde{\rho}_B^0$ & $3.7\times10^2$ & $\mathrm{kg}\,\mathrm{m}^{-3}$ \\
${\cal L}_f$ & $4\times10^8$ & $\mathrm{J}\,\mathrm{m}^{-3}$ \\
$T_m$ & 611 & K \\
$m$ & 0.25 & K$\,\mathrm{m}^{3}\,\mathrm{kg}^{-1}$ \\
$\sigma_{sl}$ & $3\times10^{-2}$ & J$\,\mathrm{m}^{-2}$ \\
$\sigma_{la}$ & $7.6\times10^{-2}$ & J$\,\mathrm{m}^{-2}$ \\
$\mu$ & $1.787\times10^{-3}$ & $\mathrm{kg}\,\mathrm{m}^{-1}\,\mathrm{s}^{-1}$ \\
\hline
\end{tabular}
\end{center}
\end{table}

From equation (\ref{asymp}), the most unstable wavenumber is given by
\begin{equation}
\alpha_{max}=\frac{1}{\sqrt{3\Gamma}},
\end{equation}
and thus we find the growth rate of the instability, $\sigma_{max}=\sigma(\alpha_{max})$ to be
\begin{equation}
\sigma_{max}=\frac{\tilde{\rho}_T^0}{\tilde{\rho}_{salt}}\frac{2}{3\sqrt{3}}\frac{1}{\sqrt{\Gamma}}.
\end{equation}
Thus for a typical film of thickness $d_0=10^{-4}$m, the wavelength of the instability will be $2.5\times10^{-5}$m with a growth rate of around 16 times the speed of growth of the crystal.

The most unstable wavenumber corresponds to a wavelength
\begin{equation}
\lambda\propto d_0\sqrt{\Gamma}=\sqrt{l_\gamma d_0},
\end{equation}
where $l_\gamma=\gamma/(\tilde{\rho}_T^0-\tilde{\rho}_B^0)$ is the capillary length corresponding the supersaturation across the film. This expresses the rule of thumb that the length scale characteristic of morphological instabilities is the geometric mean of the diffusion length and the capillary length (e.g, \cite{wors00,davi01}). It is also informative to note the dependence of $\sigma_{max}$ and $\alpha_{max}$ upon $d_0$.

In the small Peclet number limit, equation (\ref{a}) becomes
\begin{equation}
\tilde{\rho}_T^0-\tilde{\rho}_B^0\approx\mathrm{Pe}\tilde{\rho}_B^0
\end{equation}
so that the most unstable wavenumber is (dimensionally)
\begin{equation}
\tilde{\alpha}_{max}=\sqrt{\frac{\tilde{\rho}_B^0E}{3\gamma D}}
\end{equation}
and the dimensional growth rate of the instability is
\begin{equation}
\tilde{\omega}_{max}=\frac{2E}{3}\sqrt{\frac{\tilde{\rho}_B^0E}{3\gamma D}}.
\end{equation}
Therefore, as was suggested by the fact that the salt gradient in the film is independent of film thickness, {\em both} of these results are {\em independent} of $d_0$ and their sole dependence on the diffusive driving force, the evaporation rate $E$, is simply exhibited.   Although Du and Stone \cite{du96} did not measure the evaporation rate, here we estimate a value of $10^{-7}\mathrm{m}\,\mathrm{s}^{-1}$, from the typical room temperature situation in which they made their observations, and hence we find a wavelength of $23\mu$m.   

\section{Conclusion}

We have illustrated the existence of a new instability in the evaporation of impure (salty) films. Supersaturation of the film caused by evaporation at the liquid--vapor interface leads to enhanced growth of perturbations to the salt--liquid interface underlying the film.   The instability is similar to instabilities arising during the solidification of a binary alloy into a melt \cite{wors00}, however the chief difference is that supersaturation is created at some distance from the advancing salt front, whereas with the classical morphological instability of directionally solidified alloys, the supersaturation is caused by rejection of solvent immediately adjacent to the salt front. As suggested by Du and Stone \cite{du96} based on qualitative observations, this is a likely mechanism for the growth of the microscopic perturbations to the surface of growing ammonium chloride trees.  The analysis is appropriate for small Peclet number which, given typical expected room temperature evaporation rates, requires films less than O(1mm) in initial thickness. It should also be noted that solutal convection would be expected to set in for films on the order of 1mm in thickness. However we expect films present in ammonium chloride trees to be sufficiently thin to avoid these complications.

Although Du and Stone \cite{du96}  observed structures on the order of $10\mu$m, our analysis is principally intended to point out the nature of this evaporative instability and the experimental parameters of note, namely, $\gamma$ and $E$.  The wide variability in the observed scales across the surface of the film is likely to be due to gradients in evaporation rate at the interface due to local conditions and/or an underlying nonlinear bifurcation (common in diffusive type solidification problems) but the present state of the experimental evidence does not warrant a more thorough theoretical treatment.  Thus, while this mechanism provides the likely cause of the growth of microscopic perturbations as observed by Du and Stone \cite{du96}, it clearly coexists with the other phenomena that they have pointed out and we have discussed further here.  It is hoped that by focusing on a well defined situation, this analysis will provide a framework for refined experimental searches.  

\section {Acknowledgements}
\label{acknowledge}

The authors thank N.J. Balmforth and M.G. Worster for numerous helpful conversations during the evolution of this project.  This research, which began at the Geophysical Fluid Dynamics summer program at the Woods Hole Oceanographic Institution, was partially funded by National Science Foundation (NSF) grant OCE0325296, NSF grant OPP0440841 (JSW), and Department of Energy grant DE-FG02-05ER15741 (JSW).

\end{document}